\documentclass[letterpaper,twocolumn,prl,aps,superscriptaddress,floatfix]{revtex4}
\usepackage{mathptmx}

\usepackage[latin9]{inputenc}
\usepackage{amsmath}
\usepackage{amssymb}
\usepackage{graphicx}
\usepackage{esint}
\usepackage{float}
\usepackage{array}
\usepackage{ulem}
\usepackage{threeparttable}
\makeatletter

\pdfpageheight\paperheight
\pdfpagewidth\paperwidth

\providecommand{\tabularnewline}{\\}

\@ifundefined{textcolor}{}
{%
 \definecolor{BLACK}{gray}{0}
 \definecolor{WHITE}{gray}{1}
 \definecolor{RED}{rgb}{1,0,0}
 \definecolor{GREEN}{rgb}{0,1,0}
 \definecolor{BLUE}{rgb}{0,0,1}
 \definecolor{CYAN}{cmyk}{1,0,0,0}
 \definecolor{MAGENTA}{cmyk}{0,1,0,0}
 \definecolor{YELLOW}{cmyk}{0,0,1,0}
}

\usepackage{xcolor}\usepackage{soul}

\newcommand{\ket}[1]{\ensuremath{\left|#1\right\rangle}}
\definecolor{blue}{rgb}{0,0,1}
\definecolor{red}{rgb}{1,0,0}
\definecolor{green}{rgb}{0,1,0}

\usepackage[unicode=true,pdfusetitle,
 bookmarks=true,bookmarksnumbered=false,bookmarksopen=false,
 breaklinks=false,pdfborder={0 0 0},pdfborderstyle={},backref=false,colorlinks=true]
 {hyperref}
\hypersetup{
 colorlinks,linkcolor=red,citecolor=blue}
\makeatother

\begin{document}

\title{Error analysis in suppression of unwanted qubit interactions for a parametric gate in a tunable superconducting circuit}
\author{X.Y. Han}
\thanks{These two authors contributed equally to this work.}
\affiliation{Center for Quantum Information, Institute for Interdisciplinary Information Sciences, Tsinghua University, Beijing 100084, China}

\author{T.Q. Cai}
\thanks{These two authors contributed equally to this work.}
\affiliation{Center for Quantum Information, Institute for Interdisciplinary Information Sciences, Tsinghua University, Beijing 100084, China}

\author{X.G. Li}
\affiliation{Center for Quantum Information, Institute for Interdisciplinary Information Sciences, Tsinghua University, Beijing 100084, China}

\author{Y.K. Wu}
\affiliation{Center for Quantum Information, Institute for Interdisciplinary Information Sciences, Tsinghua University, Beijing 100084, China}

\author{Y.W. Ma}
\affiliation{Center for Quantum Information, Institute for Interdisciplinary Information Sciences, Tsinghua University, Beijing 100084, China}

\author{Y.L. Ma}
\affiliation{Center for Quantum Information, Institute for Interdisciplinary Information Sciences, Tsinghua University, Beijing 100084, China}

\author{J.H. Wang}
\affiliation{Center for Quantum Information, Institute for Interdisciplinary Information Sciences, Tsinghua University, Beijing 100084, China}

\author{H.Y. Zhang}
\affiliation{Center for Quantum Information, Institute for Interdisciplinary Information Sciences, Tsinghua University, Beijing 100084, China}

\author{Y.P. Song}\email{ypsong@mail.tsinghua.edu.cn}
\affiliation{Center for Quantum Information, Institute for Interdisciplinary Information Sciences, Tsinghua University, Beijing 100084, China}

\author{L.M. Duan}\email{lmduan@tsinghua.edu.cn}
\affiliation{Center for Quantum Information, Institute for Interdisciplinary Information Sciences, Tsinghua University, Beijing 100084, China}

\begin{abstract}
We experimentally demonstrate a parametric iSWAP gate in a superconducting circuit based on a tunable coupler for achieving a continuous tunability to eliminate unwanted qubit interactions. We implement the two-qubit $\rm iSWAP$ gate by applying a fast-flux bias modulation pulse on the coupler to turn on parametric exchange interaction between computational qubits. The controllable interaction can provide an extra degree of freedom to verify the optimal condition for constructing the parametric gate. Aiming to fully investigate error sources of the two-qubit gates, we perform quantum process tomography measurements and numerical simulations as varying static ZZ coupling strength. We quantitatively calculate the dynamic ZZ coupling parasitizing in two-qubit gate operation, and extract the particular gate error from the decoherence, dynamic ZZ coupling and high-order oscillation terms. Our results reveal that the main gate error comes from the decoherence, while the increase in the dynamic ZZ coupling and high-order oscillation error degrades the parametric gate performance. This approach, which has not yet been previously explored, provides a guiding principle to improve gate fidelity of parametric $\rm iSWAP$ gate by suppression of the unwanted qubit interactions. This controllable interaction, together with the parametric modulation technique, is desirable for crosstalk free multiqubit quantum circuits and quantum simulation applications.
\end{abstract}

\maketitle
\section{Introduction}
Building large superconducting circuits requires highly coherent and strongly interacting physical qubits to achieve high-fidelity gates. As the circuits become more complex, however, the fidelity of quantum algorithms will begin to be dominated by unwanted qubit interactions, increased decoherence and frequency-crowding, all inherent to traditional frequency-tuned architectures~\cite{pinto2010analysis}. Spurious unwanted qubit interactions can degrade gate performance. It thus becomes increasingly crucial to develop robust protocols for multi-qubit control~\cite{houck2012chip,gambetta2017building,Song2019}. Much efforts have been devoted to eliminating the unwanted coupling and achieving controllable interactions~\cite{mundada2019suppression,sheldon2016procedure}. Parametric scheme based on tunable couplers can help mitigate the problem of unwanted coupling and frequency crowding~\cite{gambetta2017building,mckay2016universal,Lu_2017}. In this scheme, the effective interaction between two qubits is mediated via a frequency-tunable bus, which dispersively couples to both computational qubits. To turn on the interaction between two qubits, the tunable bus is modulated by an external magnetic flux, at the qubit frequency detuning, which causes a parametric oscillating of the qubit-qubit exchange coupling and activates a resonant XX+YY interaction~\cite{mckay2016universal,reagor2018demonstration,roth2017analysis}. Such a flux-modulation scheme with microwave-only control provides frequency selectivity and allows to use fixed-frequency computational qubits, thereby minimizing the sensitvity of the qubits with respect to the sources of possible noise.

The microwave-only control gates have been proposed and realized for fixed-frequency qubits by applying one or more microwave drives~\cite{poletto2012entanglement,chow2013microwave}. In particular, a leading two-qubit gate for fixed-frequency qubits, the cross-resonance (CR) gate~\cite{chow2011simple,sheldon2016procedure}, has demonstrated fidelities greater than 99\%~\cite{sheldon2016procedure}. The coupling, however, is only effective when the qubit-qubit detuning is closely spaced compared to the anharmonicity of the qubits. Unlike drive-activated gates, the parametric exchange interaction does not decrease as the frequency detuning of qubits is larger than the anharmonicity. Therefore, it is promising for implementing entangling gate in larger circuits where a range of qubit frequencies are needed to avoid crosstalk. Although two-qubit interactions with parametric modulation have been experimentally implemented~\cite{mckay2016universal,reagor2018demonstration,roth2017analysis}, some unwanted interactions, in particular, parasitic ZZ coupling and high-order oscillation terms are omitted in the previous studies. Those unwanted interactions are always present during the two-qubit gate and, thus, degrade the gate performance. The suppression of static ZZ crosstalk has been investigated in a superconducting circuit of two qubits with a tunable two-coupler system, where the frequency of a coupler can be adjusted such that the ZZ interaction from each coupler destructively interferes~\cite{mundada2019suppression}. The crosstalk elimination for single-qubit gate has been verified via the simultaneous randomized benchmarking.  

\begin{figure}
\includegraphics{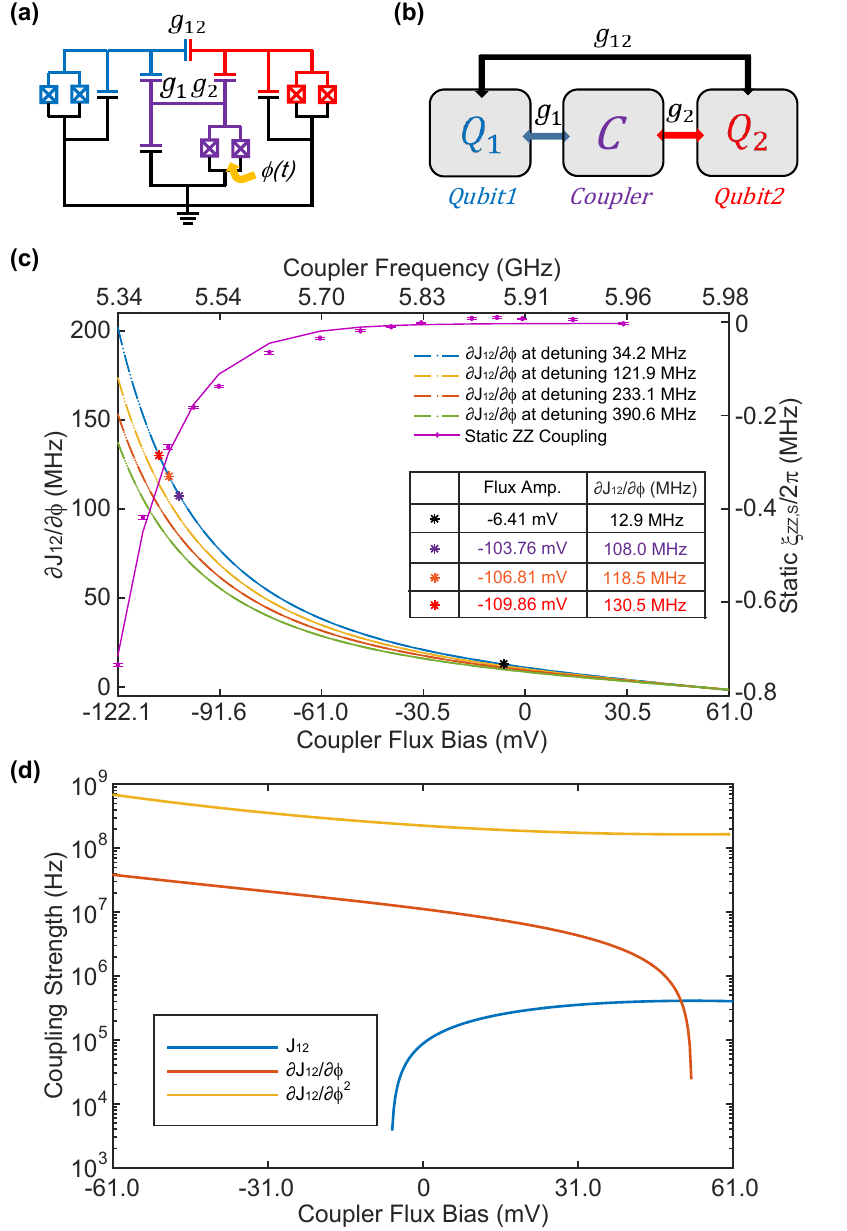}
\caption{(a) Schematic circuit of the coupler system. Parametric gate can be realized by applying a fast-flux bias $\phi (t)$ to the coupler SQUID loop. (b) Sketch of the coupler system. Two qubits are coupled directly by a coupling strength $g_{12}$ and couple to the coupler with a coupling strength $g_1$ and $g_2$, respectively. (c) Simulation curves of $\frac{\partial J_{12}}{\partial \phi}$ vs coupler flux bias (coupler frequency). We show four different qubit frequency detunings, as an example, to indicate the dependency of $\frac{\partial J_{12}}{\partial \phi}$ on the coupler flux bias (blue (first), yellow (second), red (third) and green (fourth) curves). The blue (first) curve is corresponding to the qubit frequency detuning of 34.2 MHz at which we acquire data in our experiment. The black, purple, orange and red stars superposed on the blue (first) curve represent the operating points we choose. The right panel shows the change of the static ZZ coupling strength as varying the coupler flux bias (coupler frequency). (d) Calculated data for $J_{12}$, $\frac{\partial J_{12}}{\partial \phi}$, $\frac{\partial^2 J_{12}}{\partial \phi^2}$ vs coupler flux bias. The calculation corresponds to the qubit frequency detuning of 34.2 MHz and uses the device and coupling parameters given in Table.~\ref{Table:deviceparameters} and Table.~\ref{Table:couplingparameters} in Appendix D.}
\label{fig:Fig1}
\end{figure}

In this paper, we address these crucial barriers to improving the gate fidelity by eliminating the unwanted qubit interactions. We experimentally demonstrate a parametric iSWAP gate in a superconducting circuit based on a tunable coupler for utilizing a unique competition between a positive direct and negative indirect coupling to achieve a continuous tunability~\cite{li2019tunable}. It allows a direct control of qubit interactions, with large 'on' coupling consistent with small to zero 'off' coupling, without introducing nonidealities that limit the gate performance~\cite{Bialczak_2011,chen2014qubit,geller2015tunable}. It thus can be used to more efficiently implement surface codes that require $\rm iSWAP$ gates~ \cite{ghosh2015leakage,arute2019quantum}. We employ the parametric modulation on the coupler to turn on the interaction of qubits, and aim to fully investigate the effect of the unwanted high-order oscillation terms and parasitic ZZ coupling on the two-qubit gates, which has not yet been previously explored. This controllable interaction, together with the parametric modulation technique, paves a way for crosstalk free multiqubit quantum circuits, and is desirable for scalability architectures and quantum simulation applications~\cite{lamata2018digital,kyriienko2018floquet,houck2012chip,Xu_2020,Cai_2019}.

\section{Results}
\subsection{Theory}
Our device consists of two transmon qubits ($Q_1$, $Q_2$) coupled via a frequency-tunable transmon bus (coupler). A schematic circuit is illustrated in Fig.~\ref{fig:Fig1}(a). The device parameters are similar to that presented in Ref.~\onlinecite{li2019tunable} (see Appendix C and D for the measurement setup and device parameters). The two qubits ($Q_1$, $Q_2$) each couple to the coupler ($C$) with a coupling strength  $g_i (i=1,2)$, as well as to each other through a cross-shaped capacitor with a direct coupling strength $g_{12}$, as depicted in Fig.~\ref{fig:Fig1}(b). Therefore, the two qubits interact through two channels, the direct capacitive coupling and the indirect virtual exchange coupling via the coupler. Both qubits ($\omega_{i=1,2}/2\pi= 4.9607, 4.9265$ GHz  at each sweet spot) are negatively detuned from the coupler ($\omega_{c}/2\pi=5.976$ GHz at sweet spot), $\Delta_i$($\phi$)=$\omega_i-\omega_c<0$, where $\omega_{i} (i=1,2)$ and $\omega_{c}$ are the frequencies of qubit $Q_1$, $Q_2$ and coupler, respectively. The experimentally extracted parameters, $g_{i=1,2}/2\pi=86.6, 90.6$ MHz, $g_{12}/2\pi \sim 6.74$ MHz, indicate a dispersive coupling, $g_i \ll |\Delta_{1,2}(\phi)|$. The system Hamiltonian can be written as~\cite{blais2007quantum,bravyi2011schrieffer,yan2018tunable}:
\begin{equation}\label{Eq1}
H/\hbar=\sum_{i=1,2}\frac{1}{2}{\tilde{\omega}_i}\sigma^z_i+J_{12}(\sigma^+_1\sigma^-_2+\sigma^+_2\sigma^-_1),
\end{equation}
where $\tilde{\omega_i}=\omega_i+\frac{g_i^2}{\Delta_i(\phi)}$ is the Lamb-shifted qubit frequency, $J_{12}=g_{12}+\frac{g_1 g_2}{\Delta(\phi)}$, $\Delta(\phi)=2/(\frac{1}{\Delta_1(\phi)}+\frac{1}{\Delta_2(\phi)})$. The combination of two terms, $g_{12}+\frac{g_1 g_2}{\Delta(\phi)}$, gives the total effective qubit-qubit coupling $J_{12}$, which can be adjusted by the coupler frequency through $\Delta(\phi)$. Since the tunability is continuous, one can always find a critical value $\omega_{c,off}$ to turn off the effective coupling $(J_{12}(\omega_{c,off})=0)$~\cite{niskanen2006tunable}, as well as the static ZZ coupling  $\xi_{ZZ,S}$ when two qubits are detuned in dispersive regime~\cite{li2019tunable}. 

To realize the parametric two-qubit gate, we apply a modulation flux on the coupler, $\phi(t)=\phi_{dc}+\Omega cos(\omega_{\phi}t+\varphi)$, where $\phi_{dc}$ is the DC flux bias, $\Omega cos(\omega_{\phi}t+\varphi)$ is a sinusoidal fast-flux bias modulation with an amplitude $\Omega$, frequency $\omega_{\phi}$ and phase $\varphi$. Since the coupler frequency has a nonlinear dependency on the modulation flux~\cite{koch2007charge}, a second-order DC shift of the coupler frequency results in an oscillating term at $2\omega_{\phi}$ when expanding $J_{12}$ in the parameter $\Omega cos(\omega_{\phi}t)$ to the second-order (ignoring higher-order terms)~\cite{mckay2016universal}. By substituting $J_{12}$ in Eq.~(\ref{Eq1}) with the expansion, we obtain the Hamiltonian in a rotating frame at the qubit frequencies (including the drive-induced shift):
\begin{equation}\label{Eq2}
\begin{split}
H/\hbar&=[(J_{12}+\frac{\Omega^2}{4}\frac{\partial^2 J_{12}}{\partial \phi^2})+\Omega\frac{\partial J_{12}}{\partial \phi}cos(\omega_{\phi}t)\\
&+\frac{\Omega^2}{4}\frac{\partial^2 J_{12}}{\partial \phi^2}cos(2\omega_{\phi}t)](e^{i\Delta_{12,\Omega}t}\sigma_1^{+}\sigma_2^{-}+\sigma_1^{-}\sigma_2^{+}e^{-i\Delta_{12,\Omega}t}),
\end{split}
\end{equation}
where ${\Delta_{12,\Omega}}=\Delta_{12}+\frac{\Omega^2}{4}(\frac{\partial^2 \tilde{\omega_2}}{\partial \phi^2}-\frac{\partial^2 \tilde{\omega_1}}{\partial \phi^2})$, $\Delta_{12}=\tilde{\omega_{1}}-\tilde{\omega_{2}}$. In our experiment, we apply a sinusoidal fast-flux bias modulation pulse at the frequency $\omega_{\phi}=\Delta_{12,\Omega}$ on the tunable coupler to realize the parametric $\rm iSWAP$ gate between the computational qubits~\cite{mckay2016universal}. By replacing $\Delta_{12,\Omega}$ with $\omega_{\phi}$ and Using Euler's formula we get:
\begin{equation}\label{Eq3}
\begin{split}
H/\hbar&=\frac{\Omega}{2} \frac{\partial J_{12}}{\partial \phi} (\sigma_{1}^{x} \sigma_{2}^{x}+\sigma_{1}^{y} \sigma_{2}^{y})\\
&+(J_{12}+\frac{\Omega^2}{4}\frac{\partial^2 J_{12}}{\partial \phi^2})(e^{i\omega_{\phi}t}\sigma_1^{+}\sigma_2^{-}+\sigma_1^{-}\sigma_2^{+}e^{-i\omega_{\phi}t})\\
&+\frac{\Omega^2}{8}\frac{\partial^2 J_{12}}{\partial \phi^2}(e^{-i\omega_{\phi}t}\sigma_1^{+}\sigma_2^{-}+\sigma_1^{-}\sigma_2^{+}e^{i\omega_{\phi}t})\\
&+\frac{\Omega}{2}\frac{\partial J_{12}}{\partial \phi}(e^{i2\omega_{\phi}t}\sigma_1^{+}\sigma_2^{-}+\sigma_1^{-}\sigma_2^{+}e^{-i2\omega_{\phi}t})\\
&+\frac{\Omega^2}{8}\frac{\partial^2 J_{12}}{\partial \phi^2}(e^{i3\omega_{\phi}t}\sigma_1^{+}\sigma_2^{-}+\sigma_1^{-}\sigma_2^{+}e^{-i3\omega_{\phi}t})\\
&+\cdots.
\end{split}
\end{equation}
The first term, $H_{eff}/\hbar=J_{eff}(\sigma_1^x\sigma_2^x+\sigma_1^y\sigma_2^y)$ ($J_{eff}=\frac{\Omega}{2}\frac{\partial J_{12}}{\partial \phi}$), represents the resonant exchange interaction between $Q_1$ and $Q_2$. This effective parametric modulation brings the computational qubits into resonance and can be used to implement the two-qubit gates. The others are unwanted high-order oscillation terms generated from the high-order expansion of $J_{12}$. By using our device parameters, we numerically calculate the high-order oscillation error terms $J_{12}$, $\frac{\partial J_{12}}{\partial \phi}$ and $\frac{\partial^2 J_{12}}{\partial \phi^2}$ as a function of the coupler flux bias, as shown in Fig.~\ref{fig:Fig1}(d). The coefficients of the oscillation terms are listed in Table.~\ref{Table:errordatas} in Appendix A for the four coupler flux biases used for implementing the parametric gate.

\begin{figure}
\includegraphics{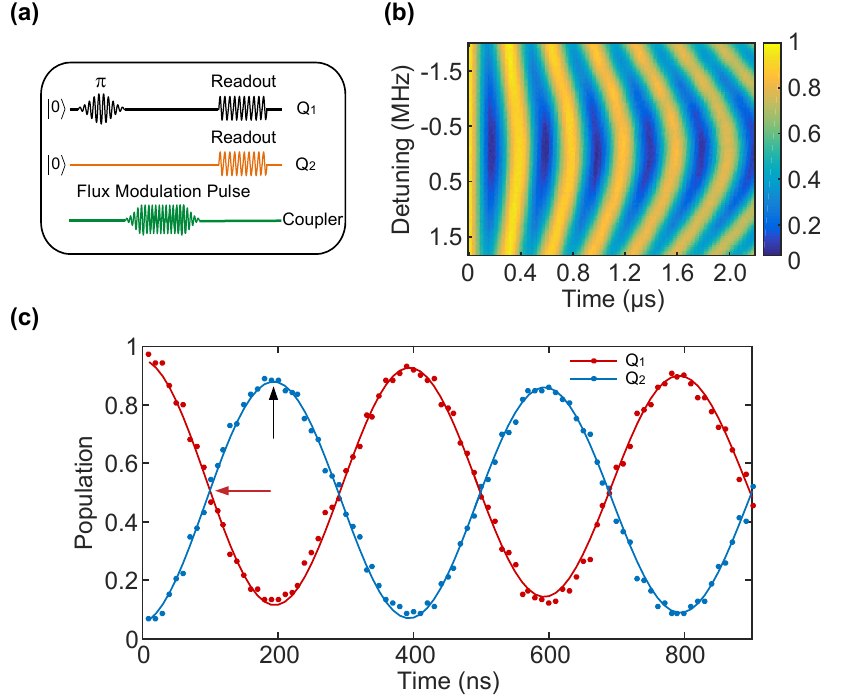}
\caption{(a) Pulse sequence for realizing the parametric $\rm iSWAP$ gate. (b) The exchange oscillation between two-qubit states \ket{10} and \ket{01} as a function of the gate length and drive frequency $\omega_{\phi}$ (with respect to the detuning between the two qubits when $\Omega=0$, $\Delta_{12,\Omega=0}=34.2$ MHz). (c) The simultaneous quantum-state population for both qubits, indicating that the excitation oscillates between the two qubits. The $\rm iSWAP$ and $\rm \sqrt{iSWAP}$ gate can be implemented at the specific evolution time $t\sim204$ ns and $t\sim102$ ns, respectively.}
\label{fig:Fig2}
\end{figure}

The coupling between higher energy levels of qubit gives rise to a cross-Kerr term $\xi_{ZZ,S}\sigma^+_1\sigma^-_1\sigma^+_2\sigma^-_2$, resulting in a static ZZ interaction~\cite{dicarlo2009demonstration,reed2013entanglement,chow2010quantum,mundada2019suppression,Kounalakis_2018}. Here, we define {$\xi_{ZZ,S}=\hbar(\omega_{11}-\omega_{01}-\omega_{10})$} as the static ZZ coupling strength in respect of performance of single-qubit gates, to distinguish it from the dynamic ZZ coupling parasitizing in two-qubit iSWAP gate operations. In contrast to the static ZZ interaction, the dynamic ZZ coupling involves time-dependent modulated components superposed on the static one in terms of the parametric drive on the coupler. Previous studies have shown that the static ZZ interaction causes dephasing in qubits and degrades gate performance~\cite{mundada2019suppression,takita2017experimental}. We measure the static ZZ interaction by a Ramsey-type measurement, which involves probing the frequency shift of $Q_1$ with initializing $Q_2$ in either its ground or excited state~\cite{reed2013entanglement,chow2010quantum}. The measurement result indicates that the static ZZ coupling strength $\xi_{ZZ,S}$ depends on the coupler frequency, as shown by a purple dot curve in Fig.~\ref{fig:Fig1}(c). At a critical coupler frequency $\omega_{c,off}=5.905$ GHz, the measured static ZZ coupling strength is $\xi_{ZZ,S} \sim 1$ kHz, close to the resolution limit for the frequency detection. We utilize simultaneous randomized benchmarking (RB) to verify the isolation of two qubits. At near zero static ZZ coupling, we get an individual RB gate fidelity of 99.44\% and 99.41\% for $Q_1$ and $Q_2$ respectively, which is nearly the same as a simultaneous RB gate fidelity of 99.45\% and 99.40\% obtained for $Q_1$ and $Q_2$. When the static ZZ coupling strength increases to $\xi_{ZZ,S}=-0.45$ MHz, however, we observe that the individual RB fidelity is almost not affected whereas the simultaneous RB gate fidelity rapidly decreases about 0.54\% for $Q_1$ and 0.93\% for $Q_2$. This result reveals that the static ZZ interaction becomes a dominant error source of the single-qubit gate operation~\cite{li2019tunable}.

\subsection{Experimental Realization of Parametric $\rm iSWAP$ Gate}
We implement two-qubit parametric $\rm iSWAP$ gate by using a pulse sequence illustrated in Fig.~\ref{fig:Fig2}(a). We initialize the two qubits in the $\ket{00}$ state with a detuning of 34.2 MHz ($\omega_{1,2}/2\pi=4.9607, 4.9265$ GHz at the sweet spot of each qubit), and set the coupler DC flux bias $\phi_{dc}\approx-6.41$ mV with respect to the coupler frequency of 5.905 GHz, at the zero static ZZ coupling strength. A $\pi$ pulse is then applied on $Q_1$. To turn on the exchange interaction between the computational qubits, a fast-flux bias modulation pulse is then applied on the coupler. Afterwards, the modulation pulse is held for a certain time to perform the $\rm iSWAP$ gate. The effective coupling strength, $J_{eff}=\frac{\Omega}{2} \frac{\partial J_{12}}{\partial \phi}$, depends on the derivative of $J_{12}$ with respect to $\phi$. The theoretical calculation of $\frac{\partial J_{12}}{\partial \phi}$ versus DC flux on the coupler is illustrated as blue (first), yellow (second), red (third) and green (fourth) curves in Fig.~\ref{fig:Fig1}(c) at four qubit frequency detunings of 34.2, 121.9, 233.1, 390.6 MHz, respectively. The effective coupling can thus be tuned from zero to a few  MHz for a moderate modulation amplitude $\Omega$. By applying the fast-flux bias modulation pulse of $\Omega cos(\omega_{\phi}t+\varphi)$ on the coupler, we measure the simultaneous quantum-state population of both qubits, as shown in Fig.~\ref{fig:Fig2}(c), indicating the excitation oscillating between the two qubits. Specific locations in Fig.~\ref{fig:Fig2}(c), $J_{eff}t=n\frac{\pi}{2}$, represent primitive two-qubit gate, $\rm iSWAP$ gate, which can be used to construct a universal gate set for quantum computing. We implement the $\rm iSWAP$ gate at a specific evolution time, indicated by a dark arrow in the figure, where $t\sim204$ ns. At the evolution time marked by a red arrow, where $t\sim102$ ns, the excitation is equally shared between both qubits, and a maximally entangled Bell state $\frac{1}{\sqrt{2}}(\ket{01}+i\ket{10})$ can be generated. Fig.~\ref{fig:Fig2}(b) shows a chevron pattern on $Q_1$ population, as a function of the coupler flux pulse length and the parametric modulation frequency $\omega_{\phi}$ with respect to the detuning between the qubits when $\Omega=0$, $\Delta_{12,\Omega=0}=34.2$ MHz. The parametric flux modulation induces a DC shift of the tunable coupler, and thus the resonance frequency of the exchange oscillating is shifted down from $\Delta_{12,\Omega=0}$ by approximately 0.25 MHz. One can expect a faster two-qubit gate by increasing the modulation strength or setting the DC flux bias to yield a lower coupler frequency but a larger value of $\frac{\partial J_{12}}{\partial \phi}$. However this will result in the increase of the leakage error and static ZZ coupling.

\begin{figure}
\includegraphics{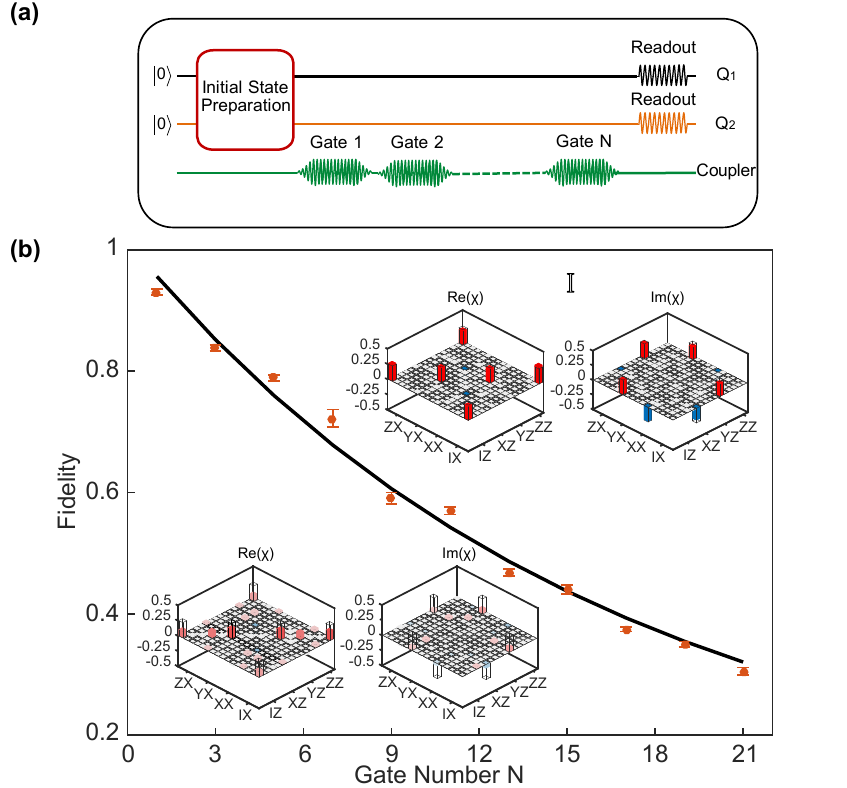}
\caption{(a) Pulse sequence for measuring an intrinsic average gate fidelity. (b) QPT fidelity decays as increasing the $\rm iSWAP$ gate number. To eliminate the measurement uncertainty, QPT is repeated five times for the measurements at each point. We fit the raw data (red dots with error bars) with a black curve based on $F=AP^N+\frac{1}{16}$ under the assumption of independent error for each gate~\cite{Li_2018,Zu_2014}. Here, $P$ is the intrinsic average gate fidelity and N is the gate number. The fitting result gives $A=0.951$ and $P=0.940$. The upper and lower insert represent the $\chi$ matrices extracted from the QPT measurements at N=1 and N=21, respectively.}
\label{fig:Fig3}
\end{figure}

We perform quantum process tomography (QPT) by implementing 16 independent two-qubit input states and construct the  matrix for $\rm iSWAP$, as shown in the upper insert of Fig.~\ref{fig:Fig3}(b). The gate fidelity can be determined from the $\chi$ matrix through the expression $F=tr(\chi_{exp} \chi_{ideal})$, where $\chi_{exp}$ and $\chi_{ideal}$ are the experimental and ideal matrix. QPT gives full information about the gate process, while it is susceptible to state preparation and measurement (SPAM) errors~\cite{chow2009randomized,Alexander2013}. To further measure the intrinsic gate error, we concatenate a series of gates and determine the fidelity decay as the number of gates (N) increases, using the pulse sequence indicated in Fig.~\ref{fig:Fig3}(a). The QPT fidelity, shown as red dots in Fig.~\ref{fig:Fig3}(b) with error bars, decays with concatenating the $\rm iSWAP$ gates. The representative $\chi$ matrices ($\chi_{exp}$ and $\chi_{ideal}$) of QPT measurements at N=1 and N=21 are shown in the upper and lower insert of Fig.~\ref{fig:Fig3}(b), respectively. By fitting the data (black curve) under the assumption of independent error for each gate~\cite{Li_2018,Zu_2014}, we obtain an average gate fidelity of 94.0\%.

\section{Discussions and Error Analyses}
\begin{figure}[h]
\includegraphics{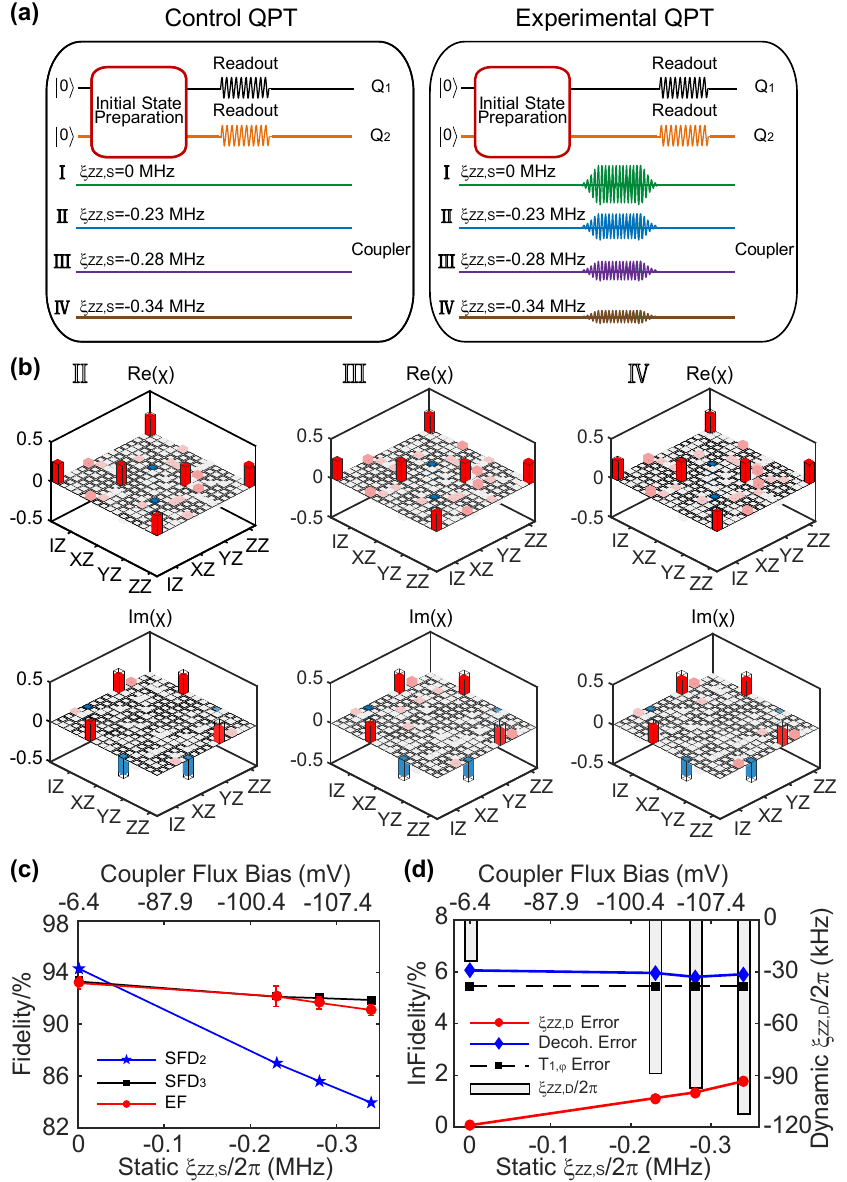}
\caption{(a) Pulse sequence for obtaining the pure $\chi$ matrix from the experimental QPT and control QPT measurements. The SPAM-error free gate fidelity can be extracted between the cases with and without the parametric modulation drive on the coupler. (b) Experimental $\chi$ matrices of the QPT measurements at the operating points with respect to the static ZZ coupling strength of $-0.23$, $-0.28$ and $-0.34$ MHz. (c) Simulation and extracted SPAM-error free gate fidelity vs static ZZ coupling strength (coupler flux bias). The black square line ($SFD_3$) represents the simulation fidelity with consideration of the qubit decoherence based on the Hamiltonian including the anharmonicity term; The blue star line ($SFD_2$) shows the simluation fidelity in consideration of qubit decoherence based on the Hamiltonian with the Hilbert space restricted to the lowest two states; The red dot line ($EF$) illustrates the SPAM-error free gate fidelity. (d) Experimental gate infidelity extracted from $\chi^{err}$ vs static ZZ coupling strength (coupler flux bias). The red dot line represents the dynamic ZZ coupling error as varying the coupler flux, while the blue diamond line shows the decoherence error. The dashed black line corresponds to the calculated gate infidelity induced by the energy relaxation and pure dephasing.The grey bars show the dynamic ZZ coupling (right panel) as varing the coupler flux (static ZZ coupling).}
\label{fig:Fig4}
\end{figure}

To characterize error sources for the parametric iSWAP gate, we perform quantum process tomography measurements as varying the DC flux bias on the coupler (static ZZ coupling strength). Here, we take a practical way to minimize the SPAM-error effect on the QPT results. Since the SPAM-error mainly results from state preparation process in our experiment, we measure the SPAM-error by conducting the process tomography with and without the two-qubit gate, and then subtract the SPAM-error from the $\chi$ matrix to obtain a pure $\chi$ matrix of the gate (see more details in Appendix B). We measure the QPT by setting the coupler at four different DC flux biases, as indicated by the black, purple, orange and red stars superposed on the blue (first) simulation curve of $\frac{\partial J_{12}}{\partial \phi}$ in Fig.~\ref{fig:Fig1}(c), with respect to the static ZZ coupling at 0, $-0.23$, $-0.28$ and $-0.34$ MHz, respectively. To eliminate the measurement uncertainty, the QPT measurement is repeated five times at each operating point. We extract the SPAM-error free gate fidelity by constructing QPT measurements, at each operating point, with (experimental QPT fidelity) and without (control QPT fidelity) the parametric modulation drive on the coupler. The corresponding pulse sequence used for the measurements is illustrated in Fig.~\ref{fig:Fig4}(a). The representative experimental matrices $\chi_{exp}$ and ideal matrices $\chi_{ideal}$ are shown as shaded bars in Fig.~\ref{fig:Fig3}(b) (upper insert) and Fig.~\ref{fig:Fig4}(b), yielding an extracted SPAM-error free gate fidelity of 93.2\%, 92.2\%, 91.7\% and 91.1\%, for the QPT data acquired at the static ZZ coupling of 0, $-0.23$, $-0.28$ and $-0.34$ MHz, respectively. We plot the extracted gate fidelity as varying the static ZZ coupling in Fig.~\ref{fig:Fig4}(c) (red dot line marked as $EF$). Apparently, the corresponding gate fidelity decreases as increasing the static ZZ coupling from zero to $-0.34$ MHz. For comparison sake, we set a same gate time of 204 ns to ensure an identical decoherence gate error at all the four operating points, by adjusting the flux modulation amplitude to compensate the change in $\frac{\partial J_{12}}{\partial \phi}$. We carefully choose experimental parameters to rule out potential error sources that may induce variation of gate fidelity as changing the DC flux bias. For instance, when the coupler frequency is reduced to be close to the computational qubit frequency, virtue excitation of coupler will become real excitation, exchanging energy between the qubit and coupler. To avoid this, we select the four DC flux biases all in the non-leakage regime (see Fig.~\ref{fig:figS4} in Appendix D). In addition, the change in the parametric modulation amplitude may attribute to the gate fidelity variation, since the large signal amplitude could lead to leakage out of the computational basis into the higher transmon levels or into the coupler. This is, however, contrary to our case. We use the moderate drive amplitude of modulation pulses, $0.115\Phi_0, 0.021\Phi_0, 0.019\Phi_0$ and $0.017\Phi_0$ ($\Phi_0$ is the flux quantum) at the four DC flux biases of the coupler. As a result, the change in the parametric gate fidelity may be  attributed to some unexcluded errors, such as the unwanted high-order oscillation terms and dynamic ZZ coupling parasitizing in the two-qubit gate operation.

\subsection{Numerical Simulations}
To verify the hypothesis above, we perform numerical simulations using the generalized Hamiltonian in Eq.~(\ref{Eq1}), restricting the Hilbert space to the lowest two states of each transmon~\cite{blais2007quantum,bravyi2011schrieffer}. We plot the simulation result in consideration of the qubit decoherence (blue star line marked as $SFD_2$) in Fig.~\ref{fig:Fig4}(c) as a function of the static ZZ coupling strength. The calculated value is slightly higher than the corresponding experimental fidelity at the zero static ZZ coupling, while the full data set demonstrates a reduction trend as the static ZZ coupling increases. At other three operating points, rapid decline of the simulation values below the experimental fidelities reflects that the model of two-level system is too simplistic and more levels need to be included for a quantitative accurate description. Consequently, we carry out more realistic simulation based on a Hamiltonian including an anharmonicity term for each qubit (see Appendix E for details). The calculated gate fidelity with considering the qubit decoherence is plotted in Fig.~\ref{fig:Fig4}(c) as a black square ($SFD_3$) line. The simulation fidelity ($SFD_3$), setting an upper bound of the two-qubit gate fidelity, again, decreases as the static ZZ coupling arises, which is in agreement with our experimental observation ($EF$). In the following, by using the error matrix method developed by A. N. Korotkovin~\cite{Alexander2013}, we quantitatively calculate the dynamic ZZ coupling strength, and extract the particular gate error from the decoherence, dynamic ZZ coupling and high-order oscillation terms. The result reveals that the main gate error comes from the decoherence, while the increase of the dynamic ZZ coupling and high-order oscillation error degrades the parametric gate performance.

\subsection{Dynamic ZZ Coupling Error}
To quantitatively verify the two-qubit gate error sources, we characterize the experimental quantum gate via an error matrix $\chi^{err}$ by separating $\chi^{err}$ and a desired unitary operation $U$ from the standard process matrix $\chi$~\cite{Alexander2013}. The standard QPT represented via $\chi$ matrix is usually defined as $\rho_{fin}=\sum_{m,n}\chi_{mn} E_m \rho_{in} E_n^{\dagger}$, where $\rho_{in}$ and $\rho_{fin}$ represent the initial and final state, the matrices $\{E_n\}$ form a basis and here we use $\{E_n\}$ = $\{I, X, Y, Z\} \otimes \{I, X, Y, Z\}$~\cite{nielsen2002quantum}. By factoring out the desired unitary operation $U=U_{\rm iSWAP}$ in $\chi$ matrix, we can obtain the $\chi^{err}$ matrix which defines the imperfections of the experimental gate. After eliminating the SPAM-error experimentally, we extract the error matrix from the $\chi$-matrix represented in the Pauli basis via the relation:
\begin{eqnarray}\label{Eq4}
\rho_{fin}=\sum_{m,n}\chi_{mn}^{err} U E_m \rho_{in} E_n^{\dagger} U^{\dagger}.
\end{eqnarray}
In an ideal case the error matrix is equal to the perfect identity operation $\chi^{I}$; otherwise, $\chi_{II,II}^{err}$, the only one large element, reflects the gate fidelity $F=tr(\chi_{exp} \chi_{ideal})=tr(\chi^{I} \chi^{err})$, and any other non-zero elements in $\chi^{err}$ indicate the imperfections of the gate operation. The imaginary parts of the elements along the left column and top row correspond to unitary imperfections, while the real parts of the elements come from decoherence error. Through the $\chi^{err}$ matrix obtained from the parametric $\rm iSWAP$ gate at the four operating points, we observe one large imaginary element $\chi_{II,ZZ}^{err} (\chi_{ZZ,II}^{err})$ which corresponds to the ZZ unitary error during the gate oepration. To determine this dynamic ZZ coupling error in the Hamiltonian, we first extract the systematic unitary gate error matrix $U^{err}=U^{actual}U^{\dagger}$ due to a slightly imperfect unitary gate operation ($U^{actual}$) according to the imaginary elements along the top row and the left column of $\chi^{err}$. We then acquire an error Hamiltonian $H^{err}$ by the expression $U^{err}=e^{iH^{err}t}$ to finally calculate the dynamic ZZ coupling term (see more details in Appendix B). The extracted dynamic ZZ coupling strength is shown as grey bars in Fig.~\ref{fig:Fig4}(d) positioned at the four operating points marked by two measures, the static ZZ and coupler flux DC bias. The average dynamic ZZ coupling strength increases accordingly, being $-23.5$, $-88.0$, $-96.4$ and $-111.0$ kHz at the four operating points, respectively. We further assess the infidelity induced by the dynamic ZZ coupling with $\Delta F_{ZZ,D}=({\rm Im}\chi_{n0}^{err})^2 / F$, where ${\rm Im}\chi_{n0}^{err}$ (or ${\rm Im}\chi_{0n}^{err})$ is the imaginary element along the top row (or left column) ($n \neq 0$); $F$ is the process fidelity. This average infidelity, valued at 0.08\%, 1.13\%, 1.35\% and 1.78\%, increases as the dynamic ZZ coupling grows which, to some extent, accounts for the decline of the fidelity at the four operating points shown in Fig.~\ref{fig:Fig4}(c) ($EF$).

\subsection{Decoherence Error}
By separating the error matrix into the coherent part and decoherence part as $\chi^{err}=\chi^{dec}+\chi^{coh}$ and diagonalizing the $\chi^{err}$ matrix, we can easily extract the value of decoherence error (see more details in Appendix B). In fact, apart from the dynamic ZZ coupling error, decoherence error constitutes the main error source in the gate operation. The decoherence error at the four operating points is depicted as a blue diamond line in Fig.~\ref{fig:Fig4}(d), demonstrating a near constant value of about 5.9\%, which is consistent with the experiment setting of the same gate time for all the four operating points. In our experiment, the most of small elements in the error matrix reflect different decoherence mechanisms. The major decoherence channel is the energy relaxation and pure dephasing of the qubits. Quantifying the decoherence process, we can simply assess the contribution to the gate infidelity due to the energy relaxation and pure dephasing mechanism, by $\Delta F_{T_{1,\phi}}=t_{gate} \cdot (\frac{1}{2T_{1}^{Q_{1}}}+\frac{1}{2T_{1}^{Q_{2}}}+\frac{1}{2T_{\phi}^{Q_{1}}}+\frac{1}{2T_{\phi}^{Q_{2}}})$, where $T_{1}^{Q_{i}}, T_{\phi}^{Q_{i}} (i=1,2)$ is the energy relaxation time and the pure dephasing time for $Q_1$ and $Q_2$, respectively; $t_{gate }=204$ ns is the gate time. Accordingly, we obtain an average induced gate infidelity of 5.4\%, as depicted as a dashed black line in Fig.~\ref{fig:Fig4}(d). The small difference between the data of dashed black line and the extracted decoherence error is due to other excluded decoherence channels, together with the interference between the decoherence error and the coherent imperfection, in particular, the coherent error from the high-order oscillation terms as discussed below. 

\subsection{High-order Oscillation Error}
In addition to the effective parametric $\rm iSWAP$ term $J_{eff}(\sigma_1^x\sigma_2^x+\sigma_1^y\sigma_2^y)$, the system Hamiltonian also involves some unwanted high-order oscillation terms at frequency of $\omega_{\phi}$, $2\omega_{\phi}$ and higher, as shown in Eq.~(\ref{Eq3}). Though the essential oscillation terms have different error accumulations from the trivial ones, yet the overall error varies and thus degrades the gate fidelity as increasing the static ZZ interaction strength (see Table.~\ref{Table:errordatas} in Appendix A and Appendix B for details). The high-order oscillation error, to the first order approximation, mainly reflects in the imaginary parts of $XX,YY,IZ,ZI$ elements in the left-column and top-row of $\chi^{err}$ matrix, which are small in our experiment. To the second-order correction, the high-order oscillation terms induce the change in the elements of $\chi_{mn}^{err}$ with $m \ne 0$ and $n \ne 0$. Although this correction is small, yet it modifies the diagonal elements, which reflects in the gate infidelity. Considering the dynamic ZZ coupling error and assuming only the energy relaxation and pure dephasing counting for the decoherence error, we estimate an upper bound of the high-order oscillation error of about 1.32\%, 1.27\%, 1.55\% and 1.72\% at the four operating points, respectively. In experiment, by increasing the qubit frequency detuning (parametric modulation frequency), we can effectively suppress this intrinsic gate error induced by the high-order oscillation terms.

\section{Conclusions}
The gate fidelity we acquired, 94.0\% (intrinsic average gate fidelity), is not very high, mainly due to the decoherence limit, but we do not think that it will become a constraint of application of our coupler scheme. In fact, we can improve the gate fidelity by fabricating the coupler with more proper parameters to acquire a larger derivative of $J_{12}$ in an appropriate operating regime, and thus significantly reduce the gate time, meanwhile maintaining a zero or near zero static ZZ coupling. 

In summary, we experimentally demonstrate a parametric iSWAP gate in a superconducting circuit based on a tunable coupler, which allows continuously varying the adjacent qubit coupling from positive to negative values. We measure the static ZZ interaction by the Ramsey-type measurement. By utilizing the simultaneous RB protocol, we observe that the static ZZ coupling degrades the single-qubit gate performance. We conduct the two-qubit $\rm iSWAP$ gate by applying the fast-flux bias modulation pulse on the coupler to turn on the parametric exchange interaction between the computational qubits. As varying the static ZZ coupling strength, we perform quantum process tomography measurements and numerical simulations to fully investigate the error sources of the parametric two-qubit gates. We quantitatively calculate the dynamic ZZ coupling and extract the particular gate error from the decoherence, dynamic ZZ coupling and high-order oscillation terms. Our results indicate that the decoherence error constitutes the main error source in the gate operation, while the increase of the dynamic ZZ coupling and high-order oscillation error degrades the parametric gate performance. This controllable interaction, together with the parametric modulation techniques, is desirable for crosstalk free multiqubit quantum circuits and quantum simulation applications.

\section*{Acknowledgments}
We thank Haiyan Wang and Chengyao Li for the technical support. We acknowledge Luyan Sun for sharing JPA fabrication parameters. This work was supported by the National Natural Science Foundation of China under Grant No.11874235, the State's Key Project of Research and Development Plan under Grant No. 2016YFA0301902 and the Tsinghua University Initiative Scientific Research Program.

\appendix

\section{Theory of parametric $\rm iSWAP$ gate}
In the Hamiltonian in Eq.~(\ref{Eq3}), the first term represents the resonant exchange interaction between $Q_1$ and $Q_2$ and its coefficient determines the exchange rate for realizing the parametric $\rm iSWAP$ gate. The other terms represent the unwanted high-order oscillation errors generated from the high-order expansion of $J_{12}$. Here we only show the expansion to $3\omega_{\phi}$. In fact, the error accumulation is attributed to all error terms~\cite{mckay2016universal,mundada2019suppression}. Although the essential oscillation terms have different error accumulations from the trivial ones, the overall errors increase as varying the DC flux bias on the coupler.

We numerically calculate the high-order oscillation error terms by using our device parameters. As shown in Fig.~\ref{fig:Fig1}(d) in the main text, we calculate $J_{12}$, $\frac{\partial J_{12}}{\partial \phi}$, $\frac{\partial^2 J_{12}}{\partial \phi^2}$ as changing the coupler flux bias. Based on the data, we then acquire the coefficients of each term in Eq.~(\ref{Eq3}) at the four different coupler flux biases discussed in the main text ($Q_1$ and $Q_2$ are biased at each sweet spot), see Table.~\ref{Table:errordatas}.

From Table.~\ref{Table:errordatas}, we can easily find that the different high-order oscillation terms show different trends when varying the coupler flux. Therefore, all these terms need to be considered to obtain the overall impact of the high-order oscillation errors, which is confirmed by our simulation result. In addition, we find that, among all the terms, the first-order oscillation term contributes most to the gate infidelity. In fact, in our experiment the gate error induced by the high-order oscillation terms shows a similar trend to the value of the first-order oscillation term as varying the coupler flux bias at the four operating points.   

\begin{table}[ht]
\begin{threeparttable}
\caption{Coefficients of oscillation terms.}
\begin{tabular}{cp{1.4cm}<{\centering}p{1.4cm}<{\centering}p{1.4cm}<{\centering}p{1.4cm}<{\centering}}
\hline
\hline
Coupler flux bias &{$J_{12}$} &{$\frac{\Omega^2}{4}\frac{\partial^2 J_{12}}{\partial \phi^2}$} &{$\frac{\Omega^2}{8}\frac{\partial^2 J_{12}}{\partial \phi^2}$} &{$\frac{\Omega}{2}\frac{\partial J_{12}}{\partial \phi}$}\tabularnewline
\hline
$-6.31$ mV &{$\approx 0$} &{$0.610$} &{0.305} &{0.645}\tabularnewline
$-103.76$ mV &{$-4.678$} &{$0.230$} &{$0.115$} &{$0.982$}\tabularnewline
$-106.81$ mV &{$-5.076$} &{$0.221$} &{$0.110$} &{$0.988$}\tabularnewline
$-109.86$ mV &{$-5.513$} &{$0.209$} &{$0.105$} &{$0.989$}\tabularnewline
\hline
\hline
\end{tabular}
\label{Table:errordatas}
\begin{tablenotes}
\item[1] Coefficients of the oscillation terms calculated by the expansion of $J_{12}$ to the second-order.
\item[2] All data are devided by $2\pi$ in unit of MHz.
\end{tablenotes}
\end{threeparttable}
\end{table}

\section{Quantum Process Tomography with Error Matrices}
To quantitatively identify the effects of the error sources during the parametric $\rm iSWAP$ gate, we take advantages of the QPT which can provide full information about the gate operation. The standard QPT is usually represented by the process matrix $\chi$ in the Pauli basis, $\{E_n\}$ = $\{I, X, Y, Z\} \otimes \{I, X, Y, Z\}$ for two qubits. However, to facilitate error analysis, we take another representation, the error matrix $\chi^{err}$ as stated in Ref.~\onlinecite{Alexander2013}, by separating the desired unitary operation  $U=U_{\rm iSWAP}$ from the standard process matrix $\chi$, see Fig.~\ref{fig:figS1}(a).  We extract the error matrix from the standard $\chi$ matrix with the relations:
\begin{equation}\label{EqS1}
\begin{split}
\chi^{err}=T \chi T^{\dagger}, T_{mn}=tr(E_{m}^{\dagger} E_{n} U^{\dagger})/d,
\end{split}
\end{equation}
where $U=U_{\rm iSWAP}$, $d=2^2$ for two-qubit parametric $\rm iSWAP$ gate. We further express the $\chi^{err}$ in matrix form as:
\begin{equation}\label{EqS2}
\begin{split}
\chi^{err}=
\left(\begin{tabular}{ccccccc}
$\chi_{00}^{err}$ & $\chi_{01}^{err}$ & $\cdots$ & $\chi_{0n}^{err}$ & $\cdots$ & $\chi_{0N}^{err}$ \\
$\chi_{10}^{err}$ & $\chi_{11}^{err}$ &  &  &  & \\
$\vdots$ &  & $\ddots$ &  &  & \\
$\chi_{m0}^{err}$ &  &  & $\chi_{mn}^{err}$ &  & \\
$\vdots$ &  &  &  & $\ddots$ & \\
$\chi_{N0}^{err}$ &  &  &  &  & $\chi_{NN}^{err}$ \\
\end{tabular}\right) \vspace{2pt}.
\end{split}
\end{equation}
In the ideal case, the error matrix is equal to the perfect identity operation $\chi^{I}$ with only one non-zero element $\chi_{00}^{err}=1$; otherwise $\chi_{00}^{err}=\chi_{II,II}^{err}$, the only one large element, reflects the gate fidelity $F=tr(\chi_{exp} \chi_{ideal})=tr(\chi^{I} \chi^{err})$, and any other non-zero elements indicate the imperfections of the gate operation. An intuitive physical picture of the $\chi^{err}-\chi^{I}$ can be depicted as the imaginary parts of the left-column and top-row elements contributing to the unitary imperfections and other elements coming from both the decoherence and the coherent errors.

\subsection{Eliminating SPAM-errors}
\begin{figure}[t]
\includegraphics{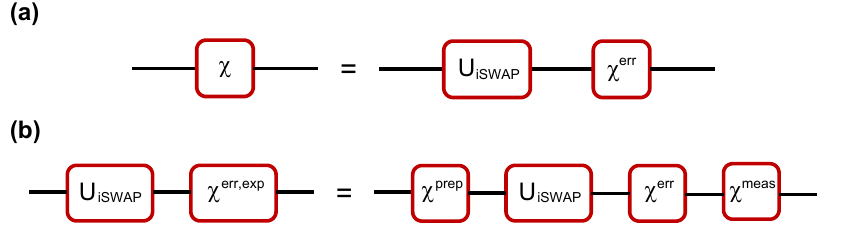}
\caption{
Circuit of quantum process tomography with error matrices. (a) Circuit diagram defining error matrix $\chi^{err}$ via its relation to $\chi$ and the desired unitary operation $U=U_{\rm iSWAP}$. (b) Circuit diagram of the SPAM-errors represented by $\chi^{prep}$ and $\chi^{meas}$.}
\label{fig:figS1}
\end{figure}

Since QPT is susceptible to state preparation and measurement errors (SPAM-errors), the error matrix $\chi^{err,exp}$ extracted from experimental $\chi$ matrix contains these SPAM-errors, see Fig.~\ref{fig:figS1}(b). Here we use $\chi^{err,exp}$ to distinguish it from $\chi^{err}$ which is simply generated from a pure gate process. Theoretically, a pure error matrix of the gate process $\chi^{err}$ can be determined by performing standard QPT of a series of gates (combinations of $X, Y, \sqrt{X}, \sqrt{Y}$ and $I$)~\cite{Alexander2013}. However, in our experiment, the SPAM-errors mainly come from the state preparation process,and thus a cumbersome identification procedure is unnecessary. The general idea is to measure SPAM-errors by conducting the standard QPT with and without the parametric $\rm iSWAP$ gate as shown in Fig.~\ref{fig:Fig4}(a) in the main text. The results are presented as $\chi^{exp}$ and $\chi^{exp,I}$ respectively. Then the error matrix of the parametric $\rm iSWAP$ gate is modified as:
\begin{equation}\label{EqS3}
\begin{split}
\chi^{err} \approx \chi^{err,exp}-(\chi^{err,exp,I}-\chi^{I}),
\end{split}
\end{equation}
where $\chi^{err,exp}=T \chi^{exp} T^{\dagger}$, $T_{mn}=tr(E_{m}^{\dagger} E_{n} U_{\rm iSWAP}^{\dagger})/d$; $\chi^{err,exp,I}=V \chi^{exp,I} V^{\dagger}$, $V_{mn}=tr(E_{m}^{\dagger} E_{n} I^{\dagger})/d$; $d=2^2$ for two-qubit parametric $\rm iSWAP$ gate.

\subsection{Extracting dynamic ZZ Coupling error, decoherence error and high-order oscillation error}
After eliminating SPAM-errors, we can simply separate the error matrix into the coherent part and decoherence part as $\chi^{err}=\chi^{dec}+\chi^{coh}$. We first extract the dynamic ZZ coupling error since this kind of systematic unitary error can be easily determined for its specific location in $\chi^{err}$ matrix . As we expand the unitary error in Pauli basis in terms of $U^{err}=\sum_{n} u_{n}^{err} E_{n}$, all the components of $U^{err}$ can be determined by the one-to-one correspondence with $u_{n}^{err}=i {\rm Im}(\chi_{n0}^{err})/F$ or $u_{n}^{err}=i {\rm Im}(\chi_{0n}^{err})/F$. The only one large imaginary element $\chi_{II,ZZ}^{err}$ ($\chi_{ZZ,II}^{err}$), on the left-column (top-row) in our experimental error matrix, indicates that the dynamic ZZ coupling contributes most to the gate infidelity while other attributions of unitary errors could be small. The relation between the unitary error evolution and error Hamiltonian can be expressed as $U^{err}=\exp^{i H^{err} t}$, where $H^{err}=\sum_{n} h_{n}^{err} E_{n}$. Therefore we can further determine $h_{ZZ}^{err}$ which corresponds to the dynamic ZZ coupling parasitizing in the parametric $\rm iSWAP$ gate by relation $h_{ZZ}^{err}=i \frac{u_{ZZ}^{err}}{t}=-\frac{{\rm Im} (\chi_{II,ZZ}^{err})}{t}$. 

Through diagonalizing the error matrix, we can get the real eigenvalues $\{ \lambda_{0}, \lambda_{1}, \cdots, \lambda_{15} \}$ of $\chi^{err}$ which are ordered as $\lambda_{0} \geq \lambda_{1} \geq \cdots \geq \lambda_{15}$. It provides an intuitive evolution representation of the gate as well as helps one to distinguish the decoherence error from the unitary imperfections~\cite{Alexander2013}. The decoherence errors at the four operating points can be extracted by the relation $\Delta F_{dec}=1-\lambda_{0}$, see Fig.~\ref{fig:Fig4}(d) in the main text. The results demonstrate a near constant value of 5.9\% which consist with the initial settings of the same gate time. From the characteristics of the error matrix, different decoherence mechanisms account for the most small peaks in the error matrix and the major decoherence channel is the energy relaxation and pure dephasing of the qubits. 

Finally, the high-order oscillation error can be simply assessed from the decoherence error and dynamic ZZ coupling error. As a certain amount of the decoherence error may be modified by its interference with the coherent imperfections, such as high-order oscillation error in our experiment, we, therefore, can estimate the upper bound of the high-order oscillation error through a relation $\Delta F_{osc}=1-F-\Delta F_{T_{1,\phi}}-\Delta F_{ZZ,D}$. The corresponding results are 1.32\%, 1.27\%, 1.55\% and 1.72\% at the four operating points, respectively.

\section{Device and Experimental Setup}
The tunable coupling device consists of two Xmon qubits ($Q_1$, $Q_2$) coupled via a coupler ($C$). Our sample is measured in a dilution refrigerator at a base temperature of about 20 mK. The device fabrication, coupler geometry and measurement circuitry are similar to that presented in Ref.~\onlinecite{li2019tunable}. For a full manipulation of the device, we use three AWGs (two Tek5014C and one Tek70002A). One AWG (Tek5014C) is connected to the input-output line for simultaneous readout of the qubits, meanwhile another one channel is used to realize individual Z control of $Q_1$ (with an extra 10 dB attenuator). The second AWG (Tek5014C), synchronized with the first one, provides two pairs of sideband modulations for XY control. The XY control signals are generated from a single microwave signal generator modulated with different sideband frequencies. This method of control guarantees stable phase differences during the quantum tomography experiments. The third AWG (Tek70002A) directly generates flux pulse to realize individual Z control of $Q_2$ and coupler. Derivative removal adiabatic gate (DRAG) pulse is used for single qubit rotation and pulse correction to reduce phase error and leakage to higher transmon levels~\cite{motzoi2009simple} . 

The readout cavity is coupled to a transmission line, which connects to a Josephson parametric amplifier(JPA)~\cite{hatridge2011dispersive,murch2013observing,Roy_2012,kamal2009signal} allowing for a high-fidelity simultaneous single-shot readout for both the qubits. The JPA, which is pumped and biased by a signal generator and a voltage source (yoko) respectively, has a gain of more than 20 dB and a bandwidth of about 313MHz at 20 mK, see gain profile in Fig.~\ref{fig:figS2}(a). It is used as the first stage of amplification followed by a HEMT amplifier at 4K and room-temperature amplifiers, allowing for high-fidelity single-shot measurements of the qubits. In the JPA circuit design, $50 \Omega$ impedance matching is applied without any other specific impedance engineering.

\section{Measurement Results}

\subsection{Readout properties}
\begin{figure}[t]
\includegraphics{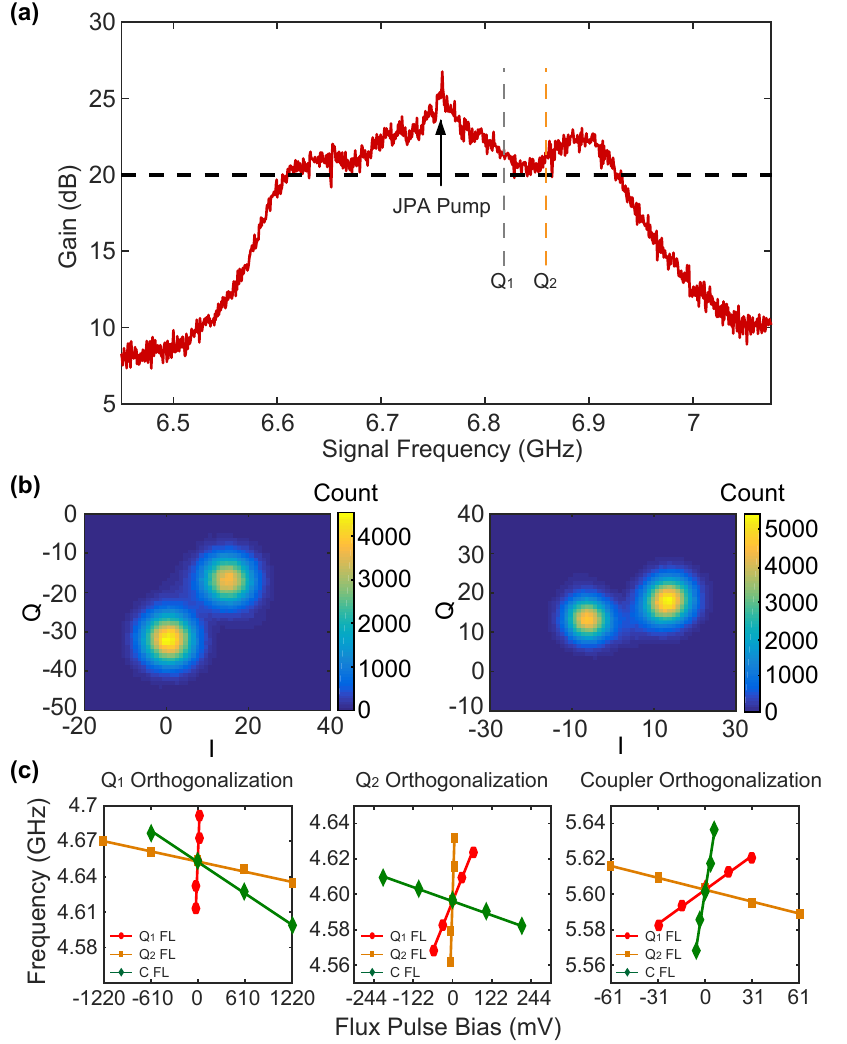}
\caption{
(a) JPA gain vs signal frequency. The JPA has a bandwidth of about 313 MHz over 20 dB gain, which is pumped seperately at 6.758 GHz, marked by a black arrow. The readout frequencies of $Q_1$ and $Q_2$ are marked by dashed grey and yellow lines respectively. (b) The I-Q data of $Q_1$ (left panel) and $Q_2$ (right panel) for the single-shot qubit-state differentiation. For each picture, the dot on the left represents qubit prepared in ground state $\ket{0}$ while the dot on the right indicates the qubit prepared in the excited state $\ket{1}$. (c) Flux bias line orthogonalization for $Q_1$ (left panel), $Q_2$ (middle panel) and the coupler (right panel) for Z matrix calibration. We measure the frequency response to each flux bias line by changing the voltage applied on it. The red dot, yellow square and green diamond lines in each panel represent the $Q_1$ flux bias line ($Q_1$ FL), $Q_2$ flux bias line ($Q_2$ FL) and coupler flux bias line ($C$ FL), respectively. The coupler frequency can be measured by the pulse sequence shown in Fig.~\ref{fig:figS3}(e). By measuring the slopes of these lines, we can get Z-crosstalk matrix which can be used to correct DC crosstalk between each flux bias line.}
\label{fig:figS2}
\end{figure}

With the help of the JPA, the single qubit gates and readout measurements are performed at the sweet spot of each qubit individually with high fidelities. The coupler frequency is tuned to 5.905 GHz to turn off the coupling between two qubits. Fig.~\ref{fig:figS2}(b) shows the I-Q data for single-shot qubit-state differentiation of each qubit when another one is in its thermal steady state. For each picture, the dot in the left represents the qubit prepared in a ground state $\ket{0}$ while the dot on the right identifies the qubit prepared in an excited state $\ket{1}$. Mismatch between the dispersive shift and decay of the readout resonator accounts for a non-perfect distinction between the ground state and excite state on each qubit. Due to the non-perfect distinction between the qubit states and thermal population of qubits and coupler, we use a calibration matrix to reconstruct the readout results based on Bayes's rule~\cite{li2019tunable}.

\subsection{Qubit parameters}
Readout frequencies, qubit frequencies, qubit coherence times, qubit anharmonicities, and dispersive shift between qubits and readout cavity are all presented in Table.~\ref{Table:deviceparameters}. The dispersive shift between each qubit and the coupler can be calculated through quantization of the qubit-coupler-qubit system. The readout frequencies of the qubits span a frequency range of about 50 MHz, well falling within the bandwidth of the JPA.

\begin{figure*}[hbt] 
\includegraphics{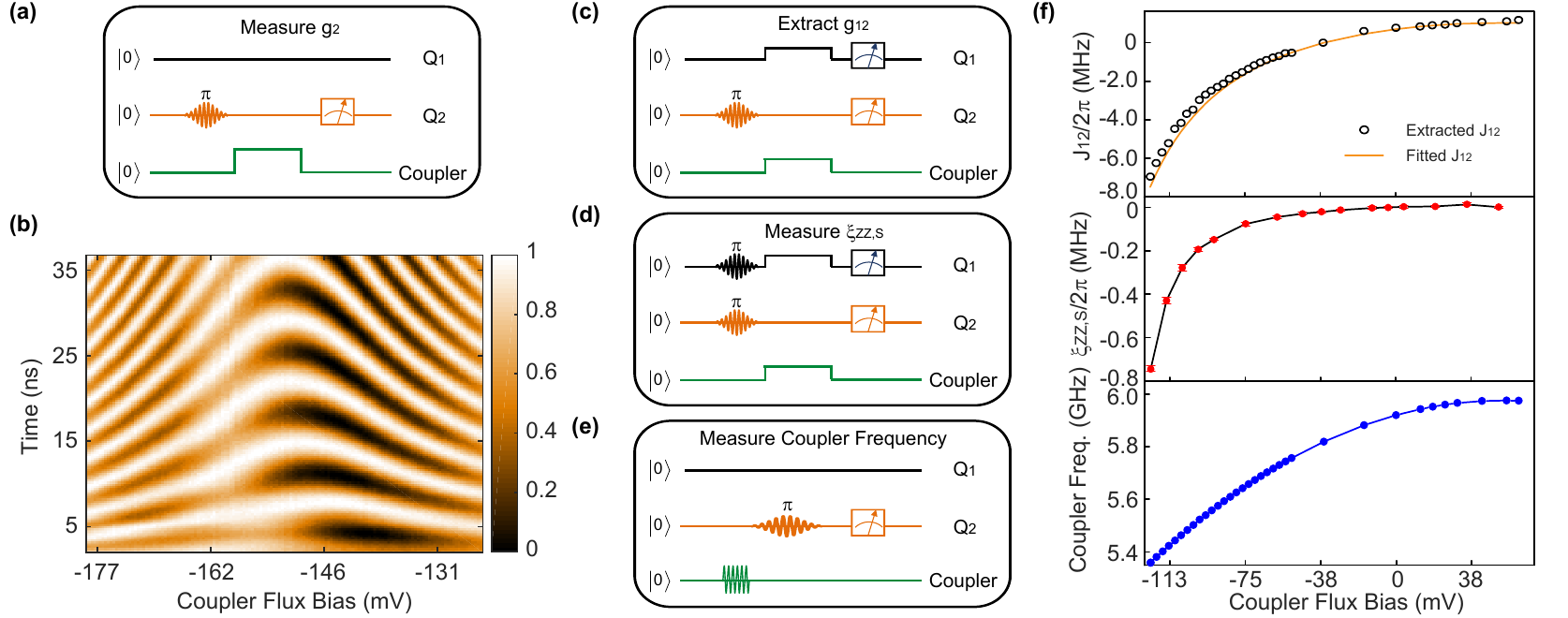}
\caption{
(a) Pulse sequence for measuring the coupling strength $g_{2}$ between $Q_2$ and the coupler. (b) Qubit-coupler energy swap. We tune the coupler frequency into resonance with one qubit, for instance, $Q_2$. At the max resonance point, we can extract the qubit-coupler direct coupling strength $g_{ic}/2\pi=76.9$ MHz (i=1,2). (c)-(e) Pulse sequence for measuring and extracting the qubit-qubit direct coupling strength $g_{12}/2\pi$, the qubit-qubit static ZZ coupling strength $\xi_{ZZ,S}/2\pi$ and the coupler frequency, respectively. (f) Top panel: Qubit-qubit effective coupling strength $J_{12}/2\pi$ vs coupler flux bias. The black dots are raw data extracted from Fig.~\ref{fig:figS4}, while the yellow line is a fitting result to extract $g_{12}$. Middle panel: Qubit-qubit static ZZ coupling strength $\xi_{ZZ,S}/2\pi$ vs coupler flux bias. The ZZ coupling strength is measured when the two qubits are biased at each sweet spot. Bottom panel: Coupler frequency vs coupler flux bias. We measure the coupler frequency by probing the dispersive shift of the qubit frequency when the coupler is pulsed into the excited state.}
\label{fig:figS3}
\end{figure*}

Although the coupler has no readout cavity, we can extract the coupler frequency by means of the dispersive shift between the qubit and coupler. The pulse sequence scheme is shown in Fig.~\ref{fig:figS3}(e). We scan the coupler frequency with a 500 ns rectangular pulse through the XY control line of $Q_1$, and then excite the $Q_2$ by a wide time-domain Gaussian pulse at its bare frequency. Once the frequency scan towards right to the coupler frequency, the coupler will be excited and dispersively shifts the frequency of $Q_2$, leading to a condition that $Q_2$ is unable to be excited with a drive pulse at its bare frequency. Therefore we can get the coupler frequency spectrum as varying the coupler flux bias, as shown in the bottom panel in Fig.~\ref{fig:figS3}(f).

\begin{table}[ht]
\caption{Device parameters.
}
\begin{tabular}{cp{1.4cm}<{\centering}p{1.4cm}<{\centering}p{1.4cm}<{\centering}}
\hline
\hline
Qubit Parameter &{$Q_1$}  &{$Q_2$} &{Coupler}\tabularnewline
\hline
Readout frequency (GHz) &{$6.825$} &{$6.864$} &{}\tabularnewline
Qubit frequency (GHz) &{$4.961$} &{$4.926$} &{$5.977$}\tabularnewline
$T_1$ (sweet point) ($\mu$s) &{$14$} &{$13.7$} &{} \tabularnewline
$T_2$ (sweet point) ($\mu$s) &{$8.4$} &{$4$} &{} \tabularnewline
$T_{2E}$ (sweet point) ($\mu$s) &{$8.7$} &{$4.4$} &{} \tabularnewline
$E_c/2\pi$ (MHz) &{$206$} &{$202$} &{$254$(Simulation)} \tabularnewline
$\chi_{qr}/2\pi$ (MHz) &{$0.4$} &{$0.4$} \tabularnewline
\hline
\hline
\end{tabular} \vspace{-6pt}
\label{Table:deviceparameters}
\end{table}

\begin{table}[ht]
\caption{Coupling parameters.
}
\begin{tabular}{cp{1.4cm}<{\centering}p{1.4cm}<{\centering}}
\hline
\hline
Coupling Parameter &{Simulation}  &{Experiment} \tabularnewline
\hline
$g_{1r}/2\pi$ (MHz) &{} &{$86.6$} \tabularnewline
$g_{2r}/2\pi$ (MHz) &{} &{$90.6$} \tabularnewline
$C_{ic} (i=1,2)$ (fF) &{$2.4$} &{} \tabularnewline
$C_{12}$ (fF)&{$0.13$} &{} \tabularnewline
$g_{i}/2\pi(i=1,2)$ (MHz) &{$81.3$} &{$76.9$} \tabularnewline
$g_{12}/2\pi$ (MHz) &{$3.8$} &{$6.74$} \tabularnewline
\hline
\hline
\end{tabular} \vspace{-6pt}
\label{Table:couplingparameters}
\end{table}

\subsection{Deconvolution and Z-crosstalk calibration matrix}
In the experiment, we use a fast-flux control to manipulate the qubits and coupler by driving a current into its corresponding SQUID loop. Nevertheless, the return path of current on each line is not explicitly controlled and, accordingly, there always exists a DC crosstalk between the each flux line. Therefore, varying the bias on any individual flux line actually changes all frequencies of the qubit and coupler. Fortunately, the frequency dependency is approximately linear for a small voltage, so the crosstalk can be corrected by the orthogonalization of the flux bias lines through multiplying the correction matrix~\cite{reed2013entanglement}. The flux bias line orthogonalization is shown in Fig.~\ref{fig:figS2}(c). We measure the frequency response from both the two qubits and the coupler, and the Z-crosstalk calibration matrix ($Q_1$ $Q_2$ $C$) is presented as follows: 
\begin{center}
$\widetilde{M}_z$=$M_z^{-1}$=
$\left(\begin{tabular}{cccc}
0.9963 & 0.0096 & 0.0264  \\
-0.0798 & 0.9997  & 0.0094   \\
-0.0116 & 0.0384  & 0.9974  \\
\end{tabular}\right)$\vspace{2pt},
\end{center} 
Where $M_z$ is the qubit frequency response matrix. Actually the Z crosstalk is very small owing to a well ground-plane connection by using airbridges~\cite{chen2014fabrication} even if the coupler is designed geometrically close to the two qubits. 

\subsection{Qubit coupling strength}
To simulate and optimize the gate fidelity for the single-qubit gates and the parametric $\rm iSWAP$ gate, we should efficiently get the qubit-qubit effective coupling strength $J_{12}$. For a full understanding and control of $J_{12}$, we need to measure and extract the qubit-coupler direct coupling strength $g_{i}$ (i=1,2) and qubit-qubit direct coupling strength $g_{12}$. First, we measure the qubit-coupler interaction by performing qubit-coupler energy-swap experiments~\cite{Bialczak_2011,ansmann2009violation}. The corresponding pulse sequence used for the measurement, $g_{2}$ as an example, is shown in Fig.~\ref{fig:figS3}(a). We tune the coupler frequency into resonance with $Q_2$, meanwhile keeping $Q_1$ away by modifying the $Q_1$ frequency. The energy-swap pattern is shown in Fig.~\ref{fig:figS3}(b). Then, we fit and extract the qubit-qubit interaction from a qubit-qubit energy-swap experiment. The pulse sequence for extracting $g_{12}$ is shown in Fig.~\ref{fig:figS3}(c). The two qubits are initialized in the ground state at each sweet spot with a detuning of 34.2 MHz. The coupler is originally set to the critical point where the coupling between the two qubits is nearly cancelled out. A $\pi$ pulse is applied on $Q_2$, followed by flux pulses on $Q_1$ and the coupler to bring $Q_1$ into resonance with $Q_2$ and turn on the qubit interaction. In this way, we can directly measure $J_{12}$ as varying the coupler flux bias when the two qubits are in resonance according to the Hamiltonian in this scheme. We extract the corresponding qubit-qubit direct coupling strength $g_{12}$ by fitting $J_{12}$ as a function of the flux bias on the coupler, as shown in the top panel of Fig.~\ref{fig:figS3}(f). All relevant parameters of the coupling strength are shown in Table.~\ref{Table:couplingparameters}.

\begin{figure}[t]
\includegraphics{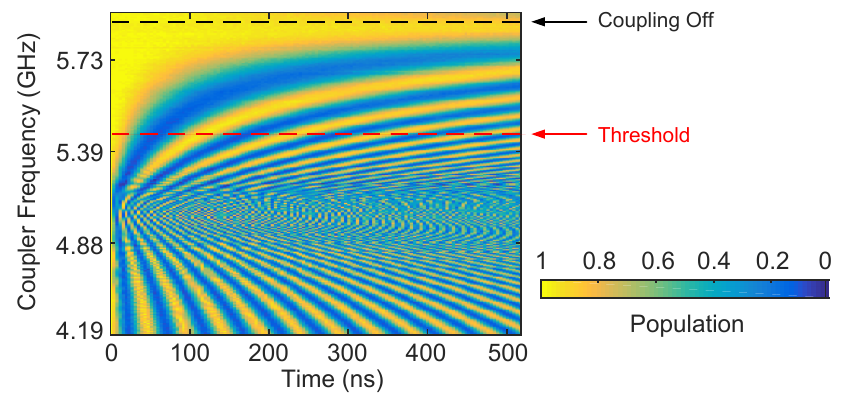}
\caption{
Population of $Q_2$ as a function of the coupler frequency and the swap time. The swap pattern clearly demonstrates the tunability of the coupling strength. The dashed dark line indicates the condition where the coupling is off, while the dashed red line marks the threshold for non-negligible excitation of the coupler. To perform a parametric $\rm iSWAP$ gate, we should choose an appropriate coupler frequency and amplitude of parametric modulation pulse to keep a non-excitation under this threshold.}
\label{fig:figS4}
\end{figure}

The single-qubit gate fidelity is affected by the qubit-qubit static ZZ-crosstalk coupling. To perform a high-fidelity single-qubit gate, the static ZZ-crosstalk should be reduced or eliminated through optimization of curcuit parameters and experimental sequences. We extract the static ZZ-crosstalk coupling strength at each sweet spot by varing the coupler frequency using Ramsey-type measurements, which involve probing the frequency of $Q_1$ with initializing $Q_2$ in either its ground or excited state~\cite{reed2013entanglement}. The pulse sequence is shown in Fig.~\ref{fig:figS3}(d). At the critical coupler frequency $\omega_{c,off}=5.905$ GHz, as shown in the middle panel of Fig.~\ref{fig:figS3}(f), the measured static ZZ coupling strength $\xi_{ZZ,S}\sim 1$ kHz, limited by the detection resolution.

To perform a fast and high fidelity two-qubit parametric $\rm iSWAP$ gate operation, we need to balance the effective coupling strength $\frac{\Omega}{2} \frac{\partial J_{12}}{\partial \phi}$ and the unwanted error terms. Apparently, a fast gate operation can be achieved by modifying the coupler frequency closer to the frequency of computational qubit, or increasing the amplitude of the coupler flux modulation pulse; however, unwanted real excitations between the qubit and coupler, or more serious high-order oscillation errors may occur. We charaterize the tunability of the coupler and estimate the threshold of operating coupler by the pulse sequence shown in Fig.~\ref{fig:figS3}(c). The results are plotted in Fig.~\ref{fig:figS4}. In Fig.~\ref{fig:figS4}, we can easily find that the qubit-qubit swap interaction is completely off in the regime marked by a dashed dark line, while, a threshold, indicated by a dashed red line, defines a regime of non-negligible excitation of the coupler. When the coupler frequency is lower than the threshold, small ripples in the population oscillation of $Q_2$ start to show up, indicating the leakage out of the computational space due to the real excitations between the qubits and coupler. Besides, the ZZ-crosstalk error may also affect the gate operation if the coupler frequency is tuned to be close to the qubit frequency. By a simulation and parameter modulation using the data shown in Fig.~\ref{fig:Fig1}(d) in the main text and Fig.~\ref{fig:figS3}(f), we can achieve the optimization of the parameters and sequences to reduce the gate error and improve the gate fidelity.

\section{Numerical Simulations of the Parametric $\rm iSWAP$ Gate}
We simulate the process of parametric $\rm iSWAP$ gate under different conditions to verify the two qubit gate error by means of Qutip~\cite{johansson2012,johansson2013}. Above all, the numerical simulations should be strictly consistent with the real experiments, so we set all the qubit parameters according to measurement results. We completely follow the experiment procedures in the simulation including pulse shape calibration, phase modulation and quantum tomography.

At first, we carry out the simulation based on the Hamiltonian which restricts the Hilbert space to the lowest two states of each transmon qubit~\cite{blais2007quantum,bravyi2011schrieffer}, The static Hamiltonian can be written as:
\begin{equation}\label{EqS4}
\begin{split}
H/\hbar&=\sum_{i=1,2}-\frac{1}{2}{\omega_i}\sigma^z_i-\frac{1}{2}{\omega_c}\sigma^z_c \\
&+\sum_{i=1,2}g_i(\sigma^+_i\sigma^-_c+\sigma^+_c\sigma^-_i)+g_{12}(\sigma^+_1\sigma^-_2+\sigma^+_2\sigma^-_1),
\end{split}
\end{equation}
where $\omega_{1,2}$ and  $\omega_{c}$ are the frequencies of $Q_1$, $Q_2$ and coupler respectively, $g_i$ (i=1,2) is the coupling of each qubit to the coupler and $g_{12}$ is the direct coupling strength between two qubits. In this model, we conduct the parametric $\rm iSWAP$ gate by applying a sinusoidal fast-flux bias modulation pulse on the coupler. We choose four different coupler flux bias points, as mentioned in the main text, to analyze the error source. After the calibration at these four operating points (the coupler frequency is 5.905, 5.491, 5.472, 5.452 GHz with respect to the static ZZ coupling at 0, $-0.23$, $-0.28$ and $-0.34$ MHz, respectively), we perform simulated quantum process tomography to characterize the gate quality. The simulation fidelities at each points are 94.3\%, 87.0\%, 85.6\% and 84.0\%. A distinct decreasing trend in fidelity with increasing the static ZZ coupling accords with the experimental results. It predicts that the unwanted high-order oscillation error and dynamic ZZ error account for degradation of the parametric $\rm iSWAP$ gate, while rapid decline of the last three points reveals that, for accurate description of the gate, we should take more energy levels into account.

Therefore, further simulation based on Hamiltonian which includes anharmonicity terms for each qubit has to be carried out. The static Hamiltonian now can be written as:
\begin{equation}\label{EqS5}
\begin{split}
H/\hbar&=\sum_{i=1,2} \omega_i a_i^\dagger a_i+ \omega_c a_c^\dagger a_c \\
&-\sum_{i=1,2}\frac{E_{c_i}}{2} a_i^\dagger a_i^\dagger a_i a_i-\frac{E_{c_c}}{2} a_c^\dagger a_c^\dagger a_c a_c \\
&+\sum_{i=1,2}g_{i}(a_i a_c^\dagger+a_c a_i^\dagger)+g_{12}(a_1 a_2^\dagger+a_2 a_1^\dagger),
\end{split}
\end{equation}
where $E_{c_i}$ (i=1,2) and $E_{c_c}$ are the anharmonicity of $Q_1$, $Q_2$ and the coupler. This model takes both the nonlinear of superconducting qubit and more relevant energy levels into account so that it is closer to the practice. All the steps remain unchanged in the simulation except for the operators when solving differential equations. We simulate the parametric gate with the qubit decoherence to explore the upper bound of the intrinsic two-qubit gate fidelity. As shown in Fig.~\ref{fig:Fig4}(c) in the main text, the simulation results and experiment data are plotted together. It is apparent that the simulation fidelity decreases as the static ZZ coupling arises, which is in agreement with our experimental observation. We attribute this dependency to the previously mentioned high-order oscillation terms as well as the dynamic ZZ coupling parasitizing in the two-qubit gate process. Through calculation of these error terms in the main text, it provides a guiding principle to improve gate fidelity of the parametric $\rm iSWAP$ gate in the future.

\normalem
%

\end{document}